\documentclass[10pt,twocolumn,letterpaper]{article}

\usepackage[pagenumbers]{cvpr} 

\usepackage{booktabs} 
\usepackage{multirow} 
\usepackage{array}    

\definecolor{cvprblue}{rgb}{0.21,0.49,0.74}
\usepackage[pagebackref,breaklinks,colorlinks,allcolors=cvprblue]{hyperref}

\def\paperID{*****} 
\def\confName{CVPR}
\def\confYear{2026}

\title{Anchoring Emotions in Text: Robust Multimodal Fusion for Mimicry Intensity Estimation}

\author{
    Lingsi Zhu$^1$ \quad Yuefeng Zou$^1$ \quad Yunxiang Zhang$^1$ \quad Naixiang Zheng$^1$ \quad Guoyuan Wang$^1$ \quad Jun Yu$^{1}$\thanks{Corresponding author.} \\
    Jiaen Liang$^2$ \quad Wei Huang$^2$ \quad Shengping Liu$^2$ \quad Ximin Zheng$^3$ \\
    $^1$University of Science and Technology of China \quad $^2$Unisound AI Technology Co., Ltd. \\
    $^3$Pingan Technology Co., Ltd.\\
    {\tt\small $^1$\{ls-zhu24, zouyuefeng.00, mesa, zhengnx, wgy2874849700\}@mail.ustc.edu.cn} \\
    {\tt\small $^1$harryjun@ustc.edu.cn} \quad {\tt\small $^2$\{liangjiaen, huangwei, liushengping\}@unisound.com} \\
    {\tt\small $^3$ZHENGXIMIN135@pingan.com.cn} 
}

\begin{document}
\maketitle

\begin{abstract}
Estimating Emotional Mimicry Intensity (EMI) in naturalistic environments is a critical yet challenging task in affective computing. The primary difficulty lies in effectively modeling the complex, nonlinear temporal dynamics across highly heterogeneous modalities, especially when physical signals are corrupted or missing. To tackle this, we propose TAEMI (Text-Anchored Emotional Mimicry Intensity estimation), a novel multimodal framework designed for the 10th ABAW Competition. Motivated by the observation that continuous visual and acoustic signals are highly susceptible to transient environmental noise, we break the traditional symmetric fusion paradigm. Instead, we leverage textual transcripts—which inherently encode a stable, time-independent semantic prior—as central anchors. Specifically, we introduce a Text-Anchored Dual Cross-Attention mechanism that utilizes these robust textual queries to actively filter out frame-level redundancies and align the noisy physical streams. Furthermore, to prevent catastrophic performance degradation caused by inevitably missing data in unconstrained real-world scenarios, we integrate Learnable Missing-Modality Tokens and a Modality Dropout strategy during training. Extensive experiments on the Hume-Vidmimic2 dataset demonstrate that TAEMI effectively captures fine-grained emotional variations and maintains robust predictive resilience under imperfect conditions. Our framework achieves a state-of-the-art mean Pearson correlation coefficient across six continuous emotional dimensions, significantly outperforming existing baseline methods.
\end{abstract}    
\section{Introduction}
\label{sec:intro}

Emotional mimicry is a fundamental mechanism in human social interaction, wherein individuals unconsciously replicate the facial expressions, vocal prosody, and body movements of others to establish empathy and strengthen social bonds. Endowing machines with the ability to accurately perceive and quantify these dynamic emotional shifts has become a critical challenge in affective computing. Consequently, the automatic estimation of Emotional Mimicry Intensity (EMI)—a task dedicated to capturing the continuous and nuanced variations in emotional transmission—has emerged as a cutting-edge research direction with profound implications for virtual assistants, social robotics, and mental health assessment.

However, existing methodologies encounter significant bottlenecks when addressing EMI estimation tasks in the wild. First, emotional mimicry is an inherently complex, multimodal temporal process. Subtle facial expressions and dynamic vocal fluctuations intertwine non-linearly over time. Current research predominantly relies on unimodal analysis or naive late-fusion strategies (e.g., simple feature concatenation), leaving the inter-modal synergy severely underutilized. For instance, baseline models utilizing ViT \cite{dosovitskiy2021imageworth16x16words} for visual features achieve a mere 0.09 mean Pearson correlation coefficient, demonstrating that unimodal visual networks struggle to capture the temporal evolution of emotional intensity. Second, fine-grained temporal alignment requires robust cross-modal semantic understanding. While acoustic features like Wav2Vec2 \cite{baevskiWav2vec20Framework2020} attain a relatively higher correlation of 0.24, they lack the semantic grounding necessary to maintain robustness in noisy, unconstrained environments. Existing fusion frameworks treat all modalities equally, failing to establish a stable semantic anchor to guide the alignment of heterogeneous emotional cues.

To address these critical challenges, we propose TAEMI (Text-Anchored Emotional Mimicry Intensity estimation), a novel multimodal framework designed specifically for the EMI track of the 10th ABAW (Affective Behavior Analysis in-the-wild) Competition \cite{zafeiriou2017aff,kollias2019deep,kollias2019face,kollias2019expression,kollias2020analysing,Kollias2025,kolliasadvancements,kollias2024distribution,kollias20246th,kollias2021affect,kollias2021analysing,kollias2021distribution,kollias2022abaw,kollias2023abaw,kollias2023abaw2,kollias2023multi,kollias20247th,kollias2024behaviour4all,kollias2025emotions,kollias2025advancements,kollias2025dvd}. The architectural design of TAEMI is driven by two core motivations:

\textbf{1) Breaking Fusion Symmetry via a Semantic Anchor:} We observe that continuous, high-framerate visual and acoustic signals are highly susceptible to transient environmental noise and temporal misalignment. In contrast, textual transcripts explicitly encode a stable, time-independent semantic context. Therefore, instead of passively concatenating features or treating all modalities symmetrically, TAEMI introduces a Text-Anchored Dual Cross-Attention mechanism. By utilizing the global textual representation as a robust query anchor, the network actively filters out frame-level redundancies and aligns the highly dynamic visual and acoustic keys around a steady semantic core.

\textbf{2) Preventing Catastrophic Degradation in the Wild:} In unconstrained real-world environments, affective signals are frequently corrupted or entirely absent (e.g., due to facial occlusions or microphone failures). A model that strictly relies on the joint presence of all modalities will suffer catastrophic performance degradation under such imperfect conditions. To mitigate this vulnerability, we integrate two synergistic robustness mechanisms. Learnable Missing-Modality Tokens provide explicit placeholder representations to maintain network stability when physical data is missing, while a Modality Dropout strategy applied during training forces the network to dynamically balance its predictive capacity and avoid over-reliance on any single modality.

We conduct systematic experiments on the Hume-Vidmimic2 dataset, which comprises 15,000 videos totaling over 25 hours of complex mimicry interactions. The experimental results validate the superiority of our approach. 

In summary, the main contributions of this work are three-fold:
\begin{itemize}
    \item We propose TAEMI, a novel multimodal architecture that shifts the fusion paradigm from naive concatenation to a Text-Anchored Dual Cross-Attention mechanism, effectively aligning visual and acoustic temporal cues using text as a semantic guide.
    \item We introduce robustness-oriented training strategies, including Learnable Missing-Modality Tokens and Modality Dropout, significantly enhancing the model's fault tolerance against incomplete real-world affective data.
    \item Our framework sets a new state-of-the-art on the Hume-Vidmimic2 dataset, achieving an overall mean Pearson correlation coefficient of 0.55 across six emotional dimensions, substantially outperforming the official baselines and existing solutions.
\end{itemize}

\section{Related Work}
\label{sec:related}

\subsection{Multimodal Emotion Recognition Techniques}
Multimodal emotion recognition fundamentally relies on fusing complementary information across visual, acoustic, and textual channels to enhance the robustness of affective analysis. Early fusion paradigms, such as the Tensor Fusion Network (TFN) \cite{zadeh2017tensorfusionnetworkmultimodal} and Dynamic Fusion Network (DFN) \cite{xu2018dynamicfusionnetworksmachine}, introduced inter-modal gating mechanisms. However, their static weight allocation limits their adaptability to the fluid temporal dynamics of emotional intensity. To address dynamic interactions, subsequent research pivoted toward advanced cross-modal modeling. For instance, the Multimodal Transformer \cite{tsaiMultimodalTransformerUnaligned2019} latently adapts streams from one modality to another across distinct time steps. Concurrently, MISA \cite{hazarikaMISAModalityInvariantSpecific2020} projects modalities into distinct invariant and specific subspaces to reduce inter-modal discrepancies, exhibiting excellent generalization capabilities across diverse sentiment analysis tasks. 

While these approaches successfully model general cross-modal interactions, they often treat all modalities symmetrically. This symmetrical treatment can exacerbate semantic misalignment when dealing with highly heterogeneous data. Compared to these existing approaches, our TAEMI framework explicitly breaks this symmetry. By designating the textual modality—which inherently carries high-level semantic priors—as a central anchor, we actively guide the temporal alignment and fusion of the more abstract visual and acoustic signals, thereby mitigating the semantic discrepancies caused by modal heterogeneity.

\subsection{Affective Computing and Emotional Mimicry Intensity}
The field of affective computing has progressively transitioned from static, categorical emotion classification to the continuous, multidimensional regression of human affective behavior in the wild. Within this scope, estimating Emotional Mimicry Intensity (EMI)—quantifying how individuals unconsciously mirror others during social interactions—has emerged as a vital task. 

Early approaches to EMI \cite{hallmen2024unimodalmultitaskfusionemotional} primarily relied on unimodal features, such as Wav2Vec 2.0 acoustic representations or Vision Transformer visual descriptors. However, recent state-of-the-art methodologies \cite{yu2024efficientfeatureextractionlate,savchenko2025hsemotionteamabaw8competition,zhang2024affectivebehaviouranalysisintegrating,yu2025technicalapproachemichallenge} underscore the absolute necessity of robust multimodal integration. Recognizing that physical acoustic and visual signals are highly susceptible to environmental noise and exhibit weak cross-modal complementarity on their own, researchers have increasingly incorporated textual modalities. Recent studies demonstrate that semantic context derived from transcribed speech (e.g., via Whisper \cite{radford2022robustspeechrecognitionlargescale_whisper}) acts as a dominant prior, significantly aiding the interpretation of complex emotional states. 

To bridge the representation gap across heterogeneous modalities, modern frameworks leverage contrastive architectures (e.g., CLIP-based alignment\cite{radford2021learningtransferablevisualmodels}) for modality-decoupled pre-training. Furthermore, dynamic fusion networks utilizing Temporal Convolutional Networks (TCN \cite{leaTemporalConvolutionalNetworks2016}) and LSTMs \cite{zhangBidirectionalLongShortTerm2015} have been proposed to adaptively allocate attention weights, striving for robustness against occlusions and noises in open-world environments. Despite these significant advancements, fine-grained temporal misalignment between continuous physical signals and discrete semantic tokens remains a critical bottleneck. Our work directly addresses this gap by introducing a text-anchored fusion pipeline paired with explicit missing-modality robustness mechanisms, ensuring that the complementary nature of these modalities is fully realized and robustly maintained.
\section{Methodology}
\label{sec:method}
\subsection{Problem Formulation}
Let $x_i=\{x_i^{a},x_i^{v},x_i^{t}\}$ denote the $i$-th multimodal input sample, where the superscripts $a, v$, and $t$ refer to the acoustic, visual, and textual  modalities, respectively. We aim to predict a 6-dimensional continuous emotion intensity vector $\hat{\mathbf{y}}_i \in \mathbb{R}^{6}$ within the real coordinate space $\mathbb{R}^{6}$, which quantifies the intensities of six target emotions: Admiration, Amusement, Determination, Empathic Pain, Excitement, and Joy. Given the corresponding ground-truth intensity vector $\mathbf{y}_i \in \mathbb{R}^{6}$, we formulate this as a multi-label regression task and optimize the model by minimizing the Mean Squared Error (MSE) between the predicted and ground-truth emotion intensities.

\begin{figure*}[htbp]
    \centering
    \includegraphics[width=\textwidth]{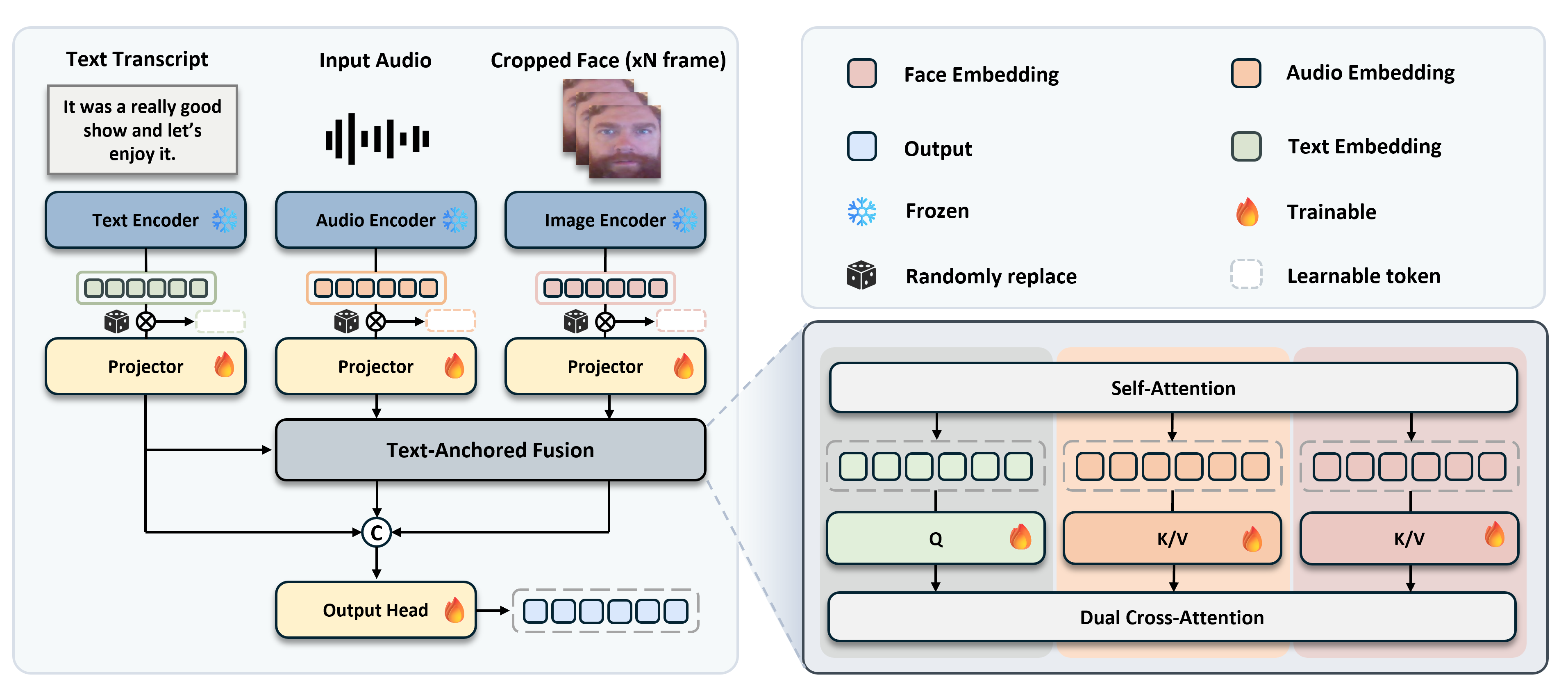}
    \caption{\textbf{The overall architecture of TAEMI.} The framework processes temporally aligned audio, visual, and textual modalities. Following modality-specific feature extraction and linear alignment, the temporal representations are fed into a Text-Anchored Dual Cross-Attention module. Here, the textual features serve as queries to aggregate pertinent audio and visual contexts. The resulting fused multimodal representation is then processed by a Multi-Layer Perceptron (MLP) to predict the continuous emotion intensity vector. During training, learnable missing-modality tokens and modality dropout are strategically incorporated to enhance the model's robustness.}
    \label{fig:pipeline}
\end{figure*}

\subsection{Data Processing and Sequence Construction}
To construct temporally aligned representations for each multimodal sample, our input pipeline processes the data as follows:

\textbf{Audio:} We truncate or zero-pad raw acoustic waveforms to a fixed duration of 12 seconds. Subsequently, a Wav2Vec 2.0 \cite{baevskiWav2vec20Framework2020} feature extractor processes these waveforms, yielding an attention mask to distinguish valid audio frames from padded regions.

\textbf{Vision:} We extract visual frame features using a pre-trained ViT \cite{vaswaniAttentionAllYou2023,dosovitskiy2021imageworth16x16words} model, clipping or padding the temporal sequence to a maximum of 400 frames.

\textbf{Text:} We adopt the Whisper \cite{radford2022robustspeechrecognitionlargescale_whisper} model to transcribe the audio. Then we encode the transcript tokens via a GTE \cite{zhang2024mgte} tokenizer, strictly constraining the maximum sequence length to 128 tokens.

Furthermore, we explicitly preserve indicators for missing modalities. During optimization, we collate variable-length sequences within batches utilizing attention masks to accurately bypass padded positions.
\subsection{Text-Anchored Dual Cross-Attention Fusion}
As illustrated in Figure \ref{fig:pipeline}, our architecture extracts and fuses features through a carefully designed text-anchored attention pipeline. We first define the initial representations for each modality.

\textbf{Audio Representation:} We define the combined audio matrix as:
$$
\mathbf{A}=\mathrm{Concat}\left(\mathbf{H}^{a},\mathbf{Z}^{a}\right)\in\mathbb{R}^{T_a\times 1027},
$$
where $\mathrm{Concat}(\cdot)$ denotes the concatenation operation, $\mathbf{H}^{a}$ represents the hidden state sequence, $\mathbf{Z}^{a}$ denotes the auxiliary logits extracted from a pre-trained emotion audio model, and $T_a$ is the temporal sequence length of the audio modality.

\textbf{Text Representation:} We extract the text feature vector as:
$$
\mathbf{t}=\mathrm{CLS}\!\left(\mathrm{GTE}(x^{t})\right)\in\mathbb{R}^{768},
$$
where $\mathrm{GTE}(\cdot)$ represents the text encoder and $\mathrm{CLS}(\cdot)$ extracts the global classification token.

\textbf{Visual Representation:} We denote the visual input as $\mathbf{V}\in\mathbb{R}^{T_v\times 768}$, where $T_v$ indicates the temporal sequence length of the visual frames.

To align these heterogeneous modalities, we first project the text, audio, and visual features into a shared latent space using simple linear layers. Following this modality alignment, we pass each modality sequence through a dedicated self-attention layer to capture intra-modal temporal dependencies. 

Subsequently, the aligned and temporally-modeled representations (denoted as $\mathbf{A}'$ and $\mathbf{V}'$ for audio and vision, respectively) enter the Dual Cross-Attention layers. The primary motivation for designating the textual modality as the central anchor lies in its temporal invariance and semantic density. Unlike continuous, high-framerate visual and acoustic signals that are highly susceptible to transient physical noise, the global textual representation $\mathbf{t}$ acts as a stable, time-independent semantic prior. Therefore, by utilizing the single-vector textual query $\mathbf{t}$ to attend over the noisy temporal keys and values of the non-text modalities, our architecture effectively filters out frame-level redundancies and aligns the heterogeneous streams around a robust semantic core:
$$
\mathbf{h}^{a}=\mathrm{CA}_{a}(\mathbf{t},\mathbf{A}'),\quad \mathbf{h}^{v}=\mathrm{CA}_{v}(\mathbf{t},\mathbf{V}'),
$$
where $\mathrm{CA}_{a}(\cdot)$ and $\mathrm{CA}_{v}(\cdot)$ denote the modality-specific cross-attention operations, yielding 512-dimensional output vectors for both $\mathbf{h}^{a}$ and $\mathbf{h}^{v}$.

We construct the fused multimodal representation by concatenating the modality-specific vectors: $\mathbf{f} = [\mathbf{t};\mathbf{h}^{a};\mathbf{h}^{v}] \in \mathbb{R}^{1792}$, where $[\cdot;\cdot]$ represents vector concatenation along the feature dimension. Finally, a Multi-Layer Perceptron (MLP) head processes this fused vector to generate the final prediction:
$$
\hat{\mathbf{y}}_i=\mathbf{W}_2\,\mathrm{Dropout}\!\left(\mathrm{GELU}(\mathbf{W}_1\mathbf{f})\right),
$$
where $\mathbf{W}_1$ and $\mathbf{W}_2$ are learnable weight matrices of the MLP, $\mathrm{GELU}(\cdot)$ is the Gaussian Error Linear Unit activation function, and $\mathrm{Dropout}(\cdot)$ is the dropout regularization operation.

\subsection{Robustness Mechanisms}
To enhance the model's resilience against missing or noisy data in real-world scenarios, we incorporate two robustness-oriented mechanisms into our architecture:

\textbf{Learnable Missing-Modality Tokens:} We inject modality-specific, trainable placeholder tokens whenever the visual or textual modality is explicitly missing from the input sample.

\textbf{Modality Dropout:} During training, we randomly replace each modality with its corresponding missing token with a probability of $p=0.1$. This strategy forces the network to avoid over-reliance on any single modality.

\subsection{Optimization Strategy}
We optimize the network using the AdamW optimizer with a learning rate of $1\times 10^{-4}$ and a weight decay of $0.01$. We employ a cosine learning rate schedule alongside a one-epoch linear warmup. Training is conducted using mixed precision, a batch size of 32, and an early stopping strategy with a patience of 5 epochs. The objective function is the Mean Squared Error (MSE) loss, defined as:
$$
\mathcal{L}_{\mathrm{MSE}}=\frac{1}{6N}\sum_{i=1}^{N}\left\|\hat{\mathbf{y}}_i-\mathbf{y}_i\right\|_2^2,
$$
where $\mathcal{L}_{\mathrm{MSE}}$ denotes the loss value, $N$ is the total number of samples in a batch, and $\|\cdot\|_2^2$ represents the squared $L_2$ norm.
\section{Experiment}
\label{sec:exp}

\subsection{Setup}  
This section details our experimental setup, encompassing the dataset, evaluation metrics, and optimization protocols.

\textbf{Datasets.} We conduct all experiments on the multimodal affective dataset Hume-Vidmimic2~\cite{Kollias2025,kollias2025dvd,kollias2025advancements,kollias2025emotions}. This dataset comprises 15,000 video samples spanning six continuous emotional dimensions: Admiration, Amusement, Determination, Empathic Pain, Excitement, and Joy. Professional annotation teams annotated each sample's emotional intensity score within the range of $[0,1]$ following standardized protocols. Table~\ref{tab:dataset} provides the statistical details of the training, validation, and test partitions.

\begin{table}[htbp]
  \centering
  \caption{\textbf{Statistical overview of the Hume-Vidmimic2 dataset partitions.} The dataset is strictly divided to ensure no data leakage between training, validation, and testing phases.}
    \begin{tabular}{ccc}
    \toprule
    Partition & Duration & Samples \\
    \midrule
    Train & 15:07:03 & 8072 \\
    Validation & 9:12:02 & 4588 \\
    Test  & 9:04:05 & 4586 \\
    \midrule
       $\sum$   & 33:23:10 & 17246 \\
    \bottomrule
    \end{tabular}%
  \label{tab:dataset}%
\end{table}%

\begin{table*}[htbp]
  \centering
  \caption{\textbf{Performance comparison on the validation set of the EMI Estimation Challenge.} We report the mean Pearson correlation coefficient ($\rho_{val}$) across six emotional targets. $V$, $A$, and $T$ denote visual, acoustic, and textual modalities, respectively. The best results are highlighted in \textbf{bold}.}
    \begin{tabular}{llcc} 
    \toprule
    Category & Model  & Modality & Mean $\rho_{val}$ \\
    \midrule
    \multirow{3}{*}{Baseline} 
    & ViT (baseline)~\cite{kollias20246th} & $V$     & 0.09 \\
    & Wav2Vec2 (baseline)~\cite{kollias20246th} & $A$     & 0.24 \\
    & ViT + Wav2Vec2 (baseline)~\cite{kollias20246th} & $V+A$     & 0.25 \\
    \midrule
    \multirow{5}{*}{\shortstack{Previous\\Solutions}}
    & Yu et al.~\cite{yu2024efficientfeatureextractionlate} & $V+A$     & 0.32 \\
    & Hallmen et al.~\cite{hallmen2024unimodalmultitaskfusionemotional} & $A$     & 0.38 \\
    & Savchenko et al.~\cite{savchenko2025hsemotionteamabaw8competition} & $V+A+T$     & 0.44 \\
    & Zhang et al.~\cite{zhang2024affectivebehaviouranalysisintegrating} & $V+A+T$     & 0.46 \\
    & Yu et al.~\cite{yu2025technicalapproachemichallenge} & $V+A+T$     & 0.51 \\
    \midrule
    \textbf{Ours} & \textbf{TAEMI} & $V+A+T$ & \textbf{0.55} \\ 
    \bottomrule
    \end{tabular}
  \label{tab:unimodal}
\end{table*}

\textbf{Evaluation.} We employ the mean Pearson correlation coefficient ($\rho$) as the primary evaluation metric for intensity estimation. It robustly measures the linear correlation between the predicted emotional response intensities and the ground-truth values. We define this metric as:  
$$
\rho = \frac{1}{6} \sum_{i=1}^{6} \rho_{i}
$$
where $\rho_{i} \ (i \in \{1,2,\dots,6\})$ denotes the correlation coefficient for each of the six emotion categories, formulated as:  
$$
\rho_{i} = \frac{\text{cov}(y_{i}, \hat{y_{i}})}{\sqrt{\text{var}(y_{i}) \text{var}(\hat{y_{i}})}} 
$$
Here, $\text{cov}(y_{i}, \hat{y_{i}})$ represents the covariance between the predicted values and the target values, while $\text{var}(y_{i})$ and $\text{var}(\hat{y_{i}})$ denote their respective variances.

\textbf{Training Details.} To ensure fair comparisons, we conduct all ablation experiments under controlled conditions, retaining the baseline optimization protocol while modifying only a single architectural factor per run. We train the model for a maximum of 50 epochs utilizing an early stopping strategy with a patience of 5 epochs. We optimize the network using the AdamW optimizer (weight decay = $0.01$) with an initial learning rate of $1\times 10^{-4}$. The learning rate scheduler follows a linear warmup (1 epoch) combined with cosine decay, updating at each optimization step. To accelerate training, we enable mixed-precision via gradient scaling and apply gradient clipping with a threshold of $100.0$. We set the batch size to 32, gradient accumulation steps to 1, and the modality dropout probability to 0.1. 

For the multimodal backbone initialization, we load the audio network from Wav2Vec 2.0~\cite{baevskiWav2vec20Framework2020,Wagner_2023}, initialize the text encoder with GTE~\cite{zhang2024mgte}, and directly utilize pre-extracted ViT representations for visual features. We execute all experiments on a single NVIDIA A800-SXM4-80GB GPU, monitoring the validation Pearson correlation coefficient ($\rho_{val}$) at each epoch.

\subsection{Main Results}
Table~\ref{tab:unimodal} summarizes the performance comparison between our proposed TAEMI architecture and existing state-of-the-art solutions on the EMI Estimation Challenge validation set. As the results indicate, relying on unimodal or bimodal inputs inherently limits the predictive capacity. By effectively integrating all three modalities ($V+A+T$) through our text-anchored fusion mechanism, TAEMI achieves a superior mean $\rho_{val}$ of \textbf{0.55}, significantly outperforming the baseline and previous top-tier solutions~\cite{yu2025technicalapproachemichallenge}.

\subsection{Ablation Study}
To validate the individual contributions of the core architectural components within TAEMI, we conduct comprehensive ablation studies. Specifically, we systematically isolate and evaluate two critical mechanisms: (1) the capability of the text-anchored dual cross-attention in capturing inter-modal dynamics, and (2) the effectiveness of our robustness modules in enhancing the model's tolerance against noisy or missing data. Table~\ref{tab:ablation} details the quantitative results of these isolated variations.

\begin{table}[htbp]
  \centering
  \caption{\textbf{Ablation study on key components of TAEMI.} We perform a leave-one-out evaluation to isolate the individual contributions of Text-Anchored Cross-Attention (TA-CA), Learnable Missing Tokens (LMT), and Modality Dropout (MD). The $\Delta$ column explicitly highlights the performance drop when a specific module is removed.}
    \begin{tabular}{ccc|cc}
    \toprule
    TA-CA & LMT & MD & Mean $\rho_{val}$ ($\uparrow$) & $\Delta$ \\
    \midrule
    \multicolumn{3}{c|}{\textit{Simple Feature Concat}} & 0.45 & -0.10 \\
    \midrule
    $\times$ & \checkmark & \checkmark & 0.48 & -0.07 \\
    \checkmark & $\times$ & \checkmark & 0.52 & -0.03 \\
    \checkmark & \checkmark & $\times$ & 0.51 & -0.04 \\
    \midrule
    \checkmark & \checkmark & \checkmark & \textbf{0.55} & - \\
    \bottomrule
    \end{tabular}%
  \label{tab:ablation}%
\end{table}%

\textbf{Effectiveness of Text-Anchored Fusion.} 
To establish a rigorous lower bound for our fusion strategy, we first evaluate a \textit{Simple Feature Concat} baseline (Table \ref{tab:ablation}, row 1). In this configuration, we completely bypass the attention pipeline—removing both the intra-modal self-attention and the text-anchored cross-attention modules. Instead, the linearly projected representations of the visual, acoustic, and textual modalities are simply flattened, concatenated along the feature dimension, and fed directly into the final MLP head. As shown in Table~\ref{tab:ablation}, this naive fusion approach yields a suboptimal performance of $0.45$. Furthermore, utilizing only intra-modal self-attention without the text-anchored cross-attention bridging them restricts the performance to $0.48$. This progression clearly validates our core motivation: since text explicitly encodes the underlying semantic context of human expressions, functioning as a stable anchor (query) to aggregate relevant visual and acoustic cues (keys/values) efficiently aligns heterogeneous emotional signals, yielding a substantial $+0.07$ performance gain over the self-attention variant.

\textbf{Impact of Robustness Mechanisms.} 
Real-world affective data frequently suffers from modality missingness. Removing the explicit Learnable Missing Tokens degrades performance to $0.52$, indicating that standard zero-padding fails to provide sufficient semantic grounding for the network when a modality is absent. More crucially, disabling the Modality Dropout ($p=0$) during training causes a noticeable drop to $0.51$. This confirms that without modality dropout, the network tends to overfit to the joint presence of all modalities, losing its predictive resilience when evaluating on imperfect validation samples. Integrating both mechanisms enables TAEMI to achieve its peak performance.

\begin{table}[htbp]
  \centering
  \caption{\textbf{Impact of different anchor modalities in the Dual Cross-Attention mechanism.} We compare utilizing Vision, Audio, or Text as the central Query to attend to the other two modalities. Text serves as the most effective query anchor due to its high-level semantic stability.}
    \begin{tabular}{lc}
    \toprule
    Anchor Modality (Query) & Mean $\rho_{val}$ ($\uparrow$) \\
    \midrule
    Vision-Anchored & 0.39 \\
    Audio-Anchored & 0.51 \\
    \textbf{Text-Anchored} & \textbf{0.55} \\
    \bottomrule
    \end{tabular}%
  \label{tab:anchor}%
\end{table}%

\textbf{Choice of Anchor Modality.} 
To rigorously justify our architectural choice of utilizing text as the central query anchor, we evaluate alternative modality arrangements. Specifically, we construct a \textit{Vision-Anchored} variant (where visual features act as the query to attend over audio and text keys/values) and an \textit{Audio-Anchored} variant (where audio features query vision and text). Table~\ref{tab:anchor} reveals that the Vision-Anchored and Audio-Anchored models achieve mean Pearson correlations of 0.49 and 0.51, respectively, both falling significantly short of our Text-Anchored approach (0.55). 

This performance gap strongly aligns with our physical intuition regarding multimodal affective data. Visual and acoustic signals are continuous, high-framerate, and highly susceptible to physical environmental noise, such as transient occlusions, lighting changes, or background overlapping sounds. Consequently, utilizing them as queries often leads to unstable attention maps. In contrast, textual transcripts represent discrete, high-level semantic priors that inherently abstract away low-level physical noise. By functioning as the query anchor, text provides a robust, noise-invariant semantic coordinate, successfully guiding the temporal alignment of the redundant physical signals without being overwhelmed by their frame-level fluctuations.
\section{Conclusion}
\label{sec:conclusion}

In this paper, we presented TAEMI, a robust and highly effective multimodal framework for continuous Emotional Mimicry Intensity (EMI) estimation. We identified that symmetric late-fusion approaches fail to bridge the semantic gap between highly heterogeneous affective signals. To overcome this, we shifted the fusion paradigm by introducing a Text-Anchored Dual Cross-Attention mechanism. By explicitly designating the textual modality as a semantic anchor, our architecture successfully guides the fine-grained temporal alignment of visual and acoustic features. Moreover, we demonstrated that incorporating Learnable Missing-Modality Tokens alongside a Modality Dropout training strategy significantly fortifies the network against the missing or noisy data frequently encountered in unconstrained environments. Evaluated on the challenging Hume-Vidmimic2 dataset for the 10th ABAW Competition, TAEMI established a new state-of-the-art with an overall mean Pearson correlation of 0.55. Moving forward, we plan to explore the integration of Large Language Models (LLMs) to extract even deeper semantic priors, further advancing the capabilities of multimodal affective computing in the wild.

\section{Acknowledgement}
This work was supported by the Natural Science Foundation of China (62276242), Hefei Municipal Natural Science Foundation (HZR2431), CAAI-MindSpore Open Fund, developed on OpenI Community.
{
    \small
    \bibliographystyle{ieeenat_fullname}
    \bibliography{main}
}


\end{document}